\documentclass[letterpaper,conference,final]{IEEEtran}

\usepackage{cite}
\usepackage{siunitx}
\DeclareSIUnit \dBm {dBm}
\DeclareSIUnit \dB {dB}
\DeclareSIUnit \Mbps {Mbps}
\DeclareSIUnit \Gbps {Gbps}
\DeclareSIUnit \mph {mph}
\usepackage{graphicx}
\usepackage{amsfonts}
\usepackage{amsmath}
\usepackage[caption=false,font=footnotesize]{subfig}
\usepackage[hyphens]{url}
\usepackage{algorithm}
\usepackage{algpseudocode}

\newcommand{\FF}[1]{{\mathbb{F}}}

\usepackage{tikz}
\usetikzlibrary{arrows}

\usepackage{enumerate}

\newcolumntype{P}[1]{>{\centering\arraybackslash}p{#1}}
\newcolumntype{M}[1]{>{\centering\arraybackslash}m{#1}}
\begin{document}
\title{Poster: High-Speed Data Dissemination over Device-to-Device Millimeter-Wave Networks for Highway Vehicular Communication}

   
\author{\IEEEauthorblockN{Andrea Tassi, Robert J. Piechocki and Andrew Nix}\\
  \IEEEauthorblockA{Department of Electrical and Electronic Engineering, University of Bristol, UK
   }}

\maketitle

\begin{abstract}
Gigabit-per-second connectivity among vehicles is expected to be a key enabling technology for sensor information sharing, in turn, resulting in safer Intelligent Transportation Systems (ITSs). Recently proposed millimeter-wave (mmWave) systems appear to be the only solution capable of meeting the data rate demand imposed by future ITS services. In this poster, we assess the performance of a mmWave device-to-device (D2D) vehicular network by investigating the impact of system and communication parameters on end-users.
\end{abstract}\vspace{-2mm}

\section{Introduction and Motivation} \label{sec.Intro}
By 2020, ten million of vehicles with onboard communication systems and a range of autonomous capabilities will be rolled out to the market. Both the European Commission's Connected-Intelligent Transportation System (C-ITS) initiative and U.S. Department of Transportation acknowledged that connectivity is pivotal in making our roads safer by enabling vehicles to exchange real-time sensor data and driving intentions. Typically, a sensor setup includes multiple proximity sensors, camcorders, and light detection and ranging (LiDAR) systems. Exchanging real-time sensor data is a challenging task that requires wireless networks providing gigabit-per-second communications links~\cite{7999286}.

Recently, millimeter-wave (mmWave) systems have been proposed as a means of overcoming the rate limitations of solutions based on LTE-A or the more traditional ITS-G5/DSRC~\cite{7999286}. Despite recent studies on device-to-device (D2D) mmWave networks~\cite{7974772,7880906}, none of these specifically refers or is directly applicable to D2D mmWave vehicular networks. With this regard, this poster focuses on the following research questions: \textbf{[Q1]} \textit{What impact do communication or system parameters (such as the antenna beamwidth or traffic intensity) have on mmWave D2D vehicular networks?} and ultimately \textbf{[Q2]} \textit{What is the maximum communication rate that can be theoretically achieved by a moving vehicle?}

\vspace{-2mm}\section{System Model and Proposed Solution} \label{sec.SM}\vspace{-1mm}
We consider a system where cars and trucks drive along a highway section consisting of $L$ parallel lanes. Assuming a system where vehicles drive on the left-hand side of the road and imposing $L$ being an even number, vehicles driving on lanes $1, \ldots, L/2$ move toward the East-to-West direction. On the other hand, lanes $L/2 + 1, \ldots, L$ are associated with the West-to-East driving direction. The $\ell$-th traffic lane is characterized by an overall traffic intensity equal to $\lambda_\ell$. We regard with $\epsilon_\ell$ the probability of a vehicle in lane $\ell$ being a truck. Thus, the density of cars and trucks in lane $\ell$ is equal to $(1-\epsilon_\ell)\lambda_\ell$ and $\epsilon_\ell\lambda_\ell$, respectively.

Since there are no restrictions preventing trucks from driving in specific traffic lanes, they can obstruct a direct link between two or more cars. In particular, whenever a truck blocks the direct link between two cars, the truck is treated as an impenetrable blockage, and no non-line of sight (NLOS) communications between the cars can occur. If there is line of sight (LOS) between two cars, transmissions will be attenuated by a path loss equal to $\ell(r) = \min\{1,C r^{-\alpha}\}$, where $C$ is the path loss intercept factor, $\alpha$ is the path loss exponent and $r$ is the distance between a transmitter and a receiver~\cite{7999286}.

\begin{figure}[t]
\centering
    \vspace{0mm}\includegraphics[width=0.4\columnwidth]{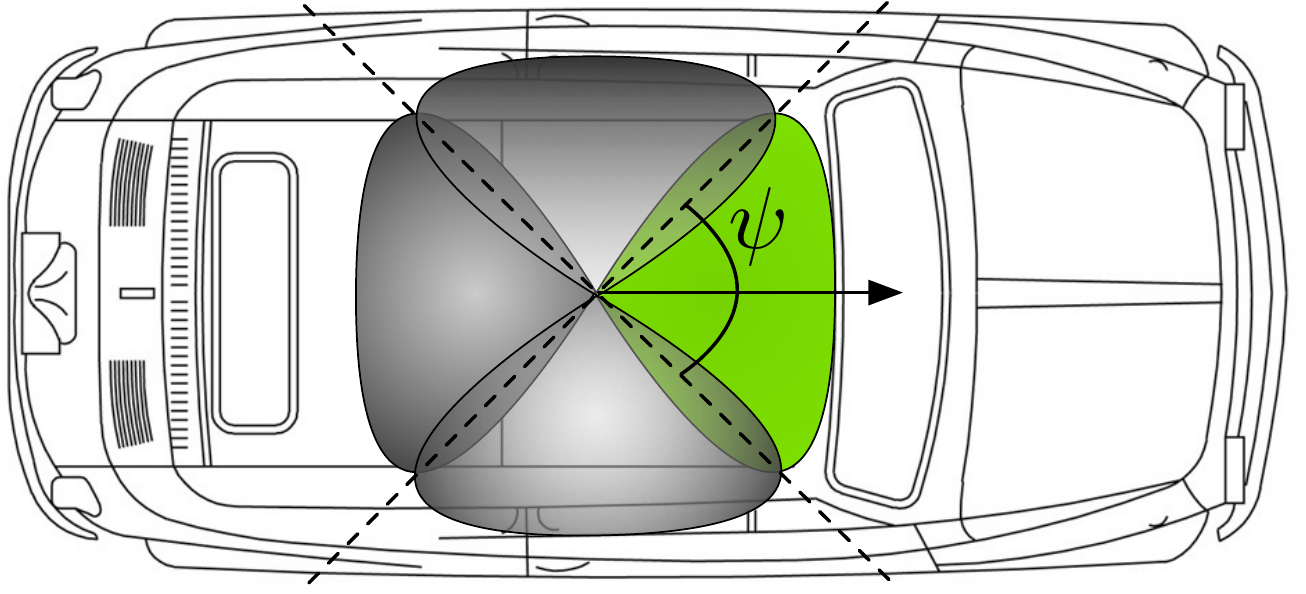}\vspace{-2mm}
\caption{Sectored approximation of the array pattern with $\psi = 90^\circ$. The picture shows the antenna boresight pointing toward the front of the vehicle.}
\label{fig.fig_SM}\vspace{-6mm}
\end{figure}

Cars are equipped with antenna arrays capable of performing directional beamforming onto the azimuth plane (see Fig.~\ref{fig.fig_SM}). To capture this feature, we follow the sectored approximation of an array pattern proposed in~\cite{7880906}. In addition, let $\psi$ be the beamwidth of the antenna main lobe, the azimuth plane is divided into $R = 2\pi/\psi$ regions, assuming $\psi$ expressed in \SI{}{\radian}). We impose that each antenna boresight can only be horizontally steered uniformly at random with steps of $\psi$ \SI{}{\radian}. As such, an antenna boresight can only point toward one the $R$ possible directions with a probability of $1/R$.

Each car can either be in receiving or transmitting mode with probability $p_\mathrm{RX}$ or $p_\mathrm{TX} = 1-p_\mathrm{RX}$, respectively. Communications among trucks have not been considered. The access to the media is regulated by a media access control (MAC) layer, which implements a time slotted system. Each slot has a duration equal to $\tau$ and it is divided into $S$ subslots, each with a duration equal to $\tau/S$. Consider a car $o$ in receiving mode. At the beginning of each slot, car $o$ (i) randomly steers its antenna beam toward one of the $R$ directions, (ii) detects all the transmitting cars $\mathcal{C}_o$ (here after referred to as \emph{transmitting cluster}) that can be received with a power not smaller than $\overline{T}$, (iii) informs each member of the transmitting cluster to transmit on a specific subslot, and (iv) puts itself in listening mode. For simplicity, we assume that $S$ is larger than the cardinality of $\mathcal{C}_o$. The signal-to-noise-plus-interference ratio (SINR) at car $o$ associated with transmissions received from the $i$-th car in $\mathcal{C}_o$ is defined as follows:
\begin{equation}
\vspace{-2mm}\!\mathrm{SINR}_o = \frac{|h_i|^2 \, \Delta_i \,\ell(r_i)}{\sigma + \mathrm{I}}, \,\,\,\,\,\text{where}\,\,\, \mathrm{I} = \sum_{j \not\in \mathcal{C}_o} |h_j|^2 \,\Delta_j\, \ell(r_j).\!\!\!\!\label{eq.sinr_o}
\end{equation}
We model the channel between any transmitting and receiving car as a Nakagami model with parameter $m$. Hence, $|h_i|^2$ follows a gamma distribution (with shape parameter $m$ and rate equal to $1$). Term $\Delta_i$ represents the overall transmitting and receiving antenna gains and $\sigma$ is the thermal noise power normalized with respect to the transmission power $P_t$. All the transmissions from those cars, which do not belong to $\mathcal{C}_o$ are regarded as interference, which has power $\mathrm{I}$.

\section{Numerical Results and Discussion} \label{sec.Res}\vspace{-1mm}
\begin{table}[t]
\centering
\caption{Main simulation parameters.}
\label{tab.sim}
{\scriptsize
\begin{tabular}{|P{2.3cm}|P{5.8cm}|}
\hline \textbf{Parameter}  & \textbf{Value}\\
\hline Road length, L, Lane width  & \SI{20}{\kilo\meter}, 4, \SI{3.7}{\meter}\\
\hline \vspace{-1.6mm}$\lambda_\ell$ & \vspace{-1.6mm}$\{10, \ldots, 60\} \cdot 10^{-3}$\\
\hline \vspace{-1.6mm}$\epsilon_1, \ldots, \epsilon_4$ & \vspace{-1.6mm}$[0.1, 0.05, 0.05, 0.1]$\\
\hline \vspace{0mm}Mobility model & \vspace{-2mm}Krauss car-following mobility model~\cite{7999286}; maximum vehicle speed equal to \SI{96}{\kilo\meter\per\hour} (trucks), \SI{112}{\kilo\meter\per\hour} (cars). \\
\hline \vspace{-2mm}Vehicle dimensions & \vspace{-2mm}$\SI{11.2}{\meter}\times\SI{2.52}{\meter}$ (trucks); $\SI{4}{\meter}\times\SI{2.52}{\meter}$ (cars);\\
\hline \vspace{-1.6mm}Carrier frequency, $W$ & \vspace{-1.6mm}$\SI{28}{\giga\hertz}$, \SI{2.16}{\giga\hertz}\\
\hline \vspace{-1.6mm}$\mathrm{C}$, $\alpha$ & \vspace{-1.6mm}Free space path loss at \SI{1}{\meter}, $2.6$~\cite{7999286}\\
\hline \vspace{-1.6mm}$p_\mathrm{RX}, p_\mathrm{TX}$ & \vspace{-1.6mm}$0.5$\\
\hline $\overline{T}$, $m$, $P_t$ & Set to allow a $\SI{100}{\meter}$ coverage, $3$, \SI{1}{\watt}~\cite{7999286}\\
\hline
\end{tabular}
}
\end{table}

Through Monte Carlo simulations having their main parameters set as reported in Table~\ref{tab.sim}, we estimate: (i) the SINR outage probability $\mathrm{P_T}(\theta) = \mathbb{P}(\mathrm{SINR}_o < \theta)$ of car $o$ for a threshold $\theta$, and (ii) the rate coverage probability $\mathrm{R_C}(\kappa)$ of car $o$ for a threshold $\kappa$, defined as the probability that car $o$ experiences a reception rate not smaller than $\kappa$. In this performance investigation, we focus on cars (in receiving mode) driving on lane $2$ as their communications will be affected by a stronger interference component compared to the users circulating on an outermost lane.

\begin{figure}[t]
\centering
    \includegraphics[width=0.95\columnwidth]{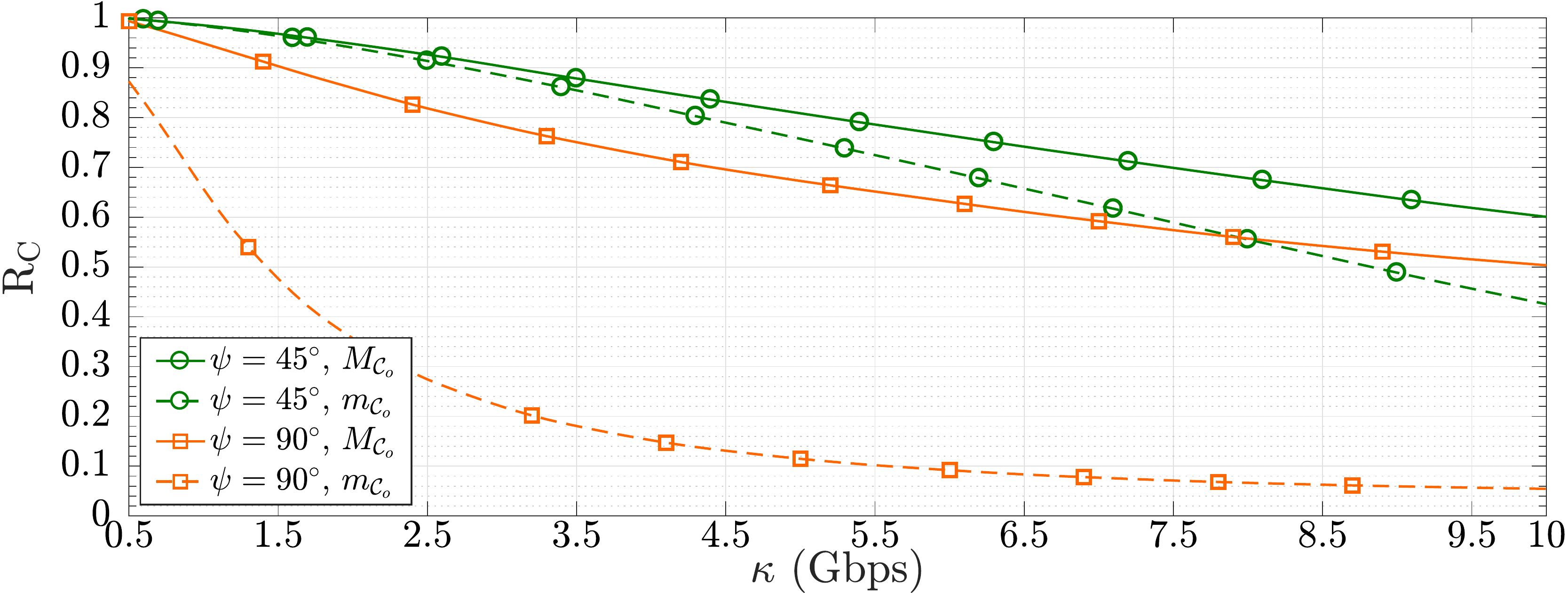}\vspace{-5mm}
\caption{Rate coverage probability $\mathrm{R_C}$ associated with $M_{\mathcal{C}_o}$ and $m_{\mathcal{C}_o}$, as a function of $\kappa$, for $(1-\epsilon_2)\lambda_2 = 5.7 \cdot 10^{-2}$ and $\psi = \{45^\circ, 90^\circ\}$.}
\label{fig.f_2}
\end{figure}

Fig.~\ref{fig.f_2} shows $\mathrm{R_C}$ as a function of $\kappa$ for a traffic density equal to $60$ cars per-kilometer ($\lambda_\ell = 6 \cdot 10^{-2}$, for $\ell = 1, \ldots, L$), for $\psi$ equal to $45^\circ$ and $90^\circ$. The figure also compares the rate coverage probability of transmissions originating from user $M_{\mathcal{C}_o}$ and $m_{\mathcal{C}_o}\in\mathcal{C}_o$ determining the maximum and minimum value of $\mathrm{SINR}_o$, respectively. From Fig.~\ref{fig.f_2}, it follows that the more the beamwidth reduces, the more the rate that user $o$ can expect to achieve from the transmitting cluster will increase. Furthermore, the figure suggests that decreasing $\psi$, significantly reduces the performance gap between $M_{\mathcal{C}_o}$ and $m_{\mathcal{C}_o}$ -- thus reducing the performance heterogeneity in $\mathcal{C}_o$.

\begin{figure}[t]
\centering
    \vspace{0.007in}\includegraphics[width=0.95\columnwidth]{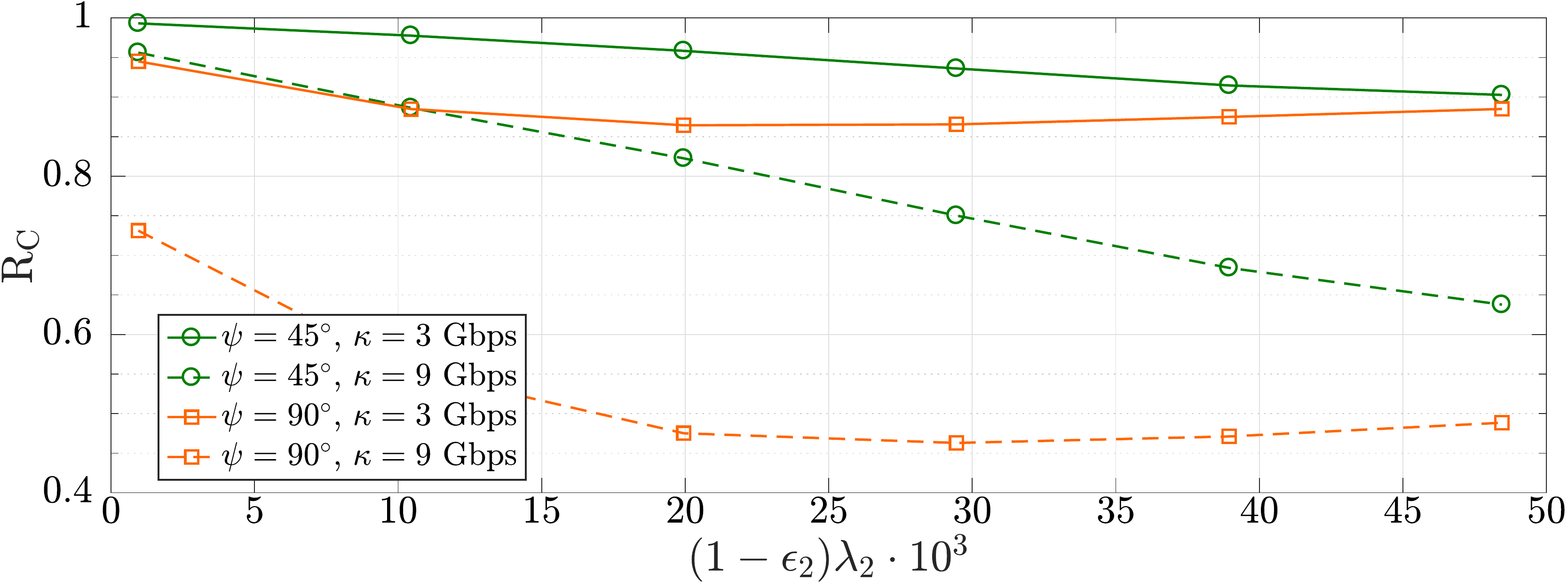}
\caption{Rate coverage probability $\mathrm{R_C}$ associated with $M_{\mathcal{C}_o}$ as a function of $(1-\epsilon_2)\lambda_2$, for $\psi = \{45^\circ, 90^\circ\}$.}
\label{fig.f_3}
\end{figure}

\begin{figure}[t]
\centering
    \includegraphics[width=0.95\columnwidth]{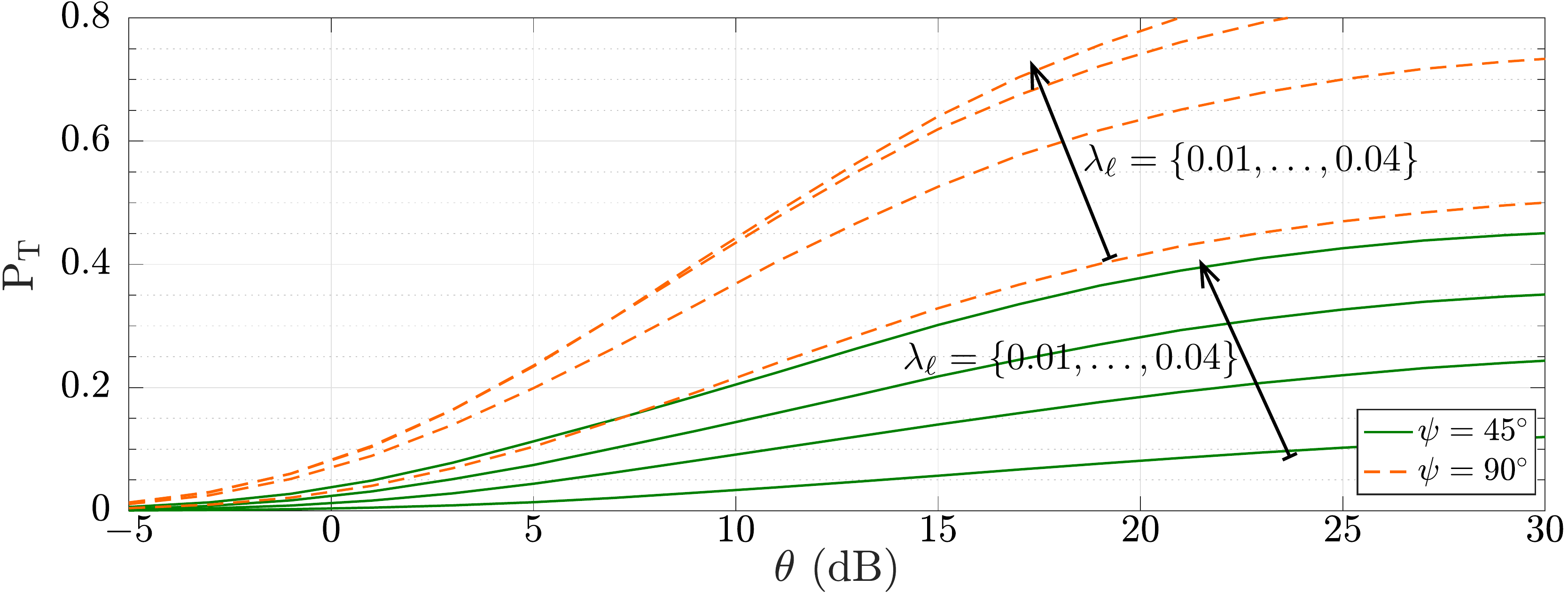}\vspace{-4mm}
\caption{SINR outage probability $\mathrm{P_T}$ associated with $M_{\mathcal{C}_o}$ as a function of $\theta$, for $\lambda_2 = \{0.01, \ldots, 0.04\}$ and $\psi = \{45^\circ, 90^\circ\}$.}\vspace{-5mm}
\label{fig.f_4}
\end{figure}

Fig.~\ref{fig.f_3} shows the rate coverage probability associated with user $M_{\mathcal{C}_o}$ as a function of the car density on lane $2$, for $\kappa$ equal to $\SI{3}{\Gbps}$ and $\SI{9}{\Gbps}$. We observe that $\mathrm{R_C}$ increases as the beamwidth, the value of $\kappa$ or the vehicle density decrease. These conclusions are further reinforced by Fig.~\ref{fig.f_4} showing that the SINR coverage probability for different values of $\lambda_2$.

\section{Conclusions}\vspace{0mm}
The adoption of a slotted communication system determines that cars belonging to the same transmitting cluster do not interfere among themselves while communicating to user $o$. Despite this, the overall interference contribution is not negligible. In particular, with regards to \textbf{[Q1]}, for a fixed antenna beamwidth, the rate coverage probability and hence, the SINR outage probability are substantially impacted by the density of vehicles. In addition, the smaller the value of $\psi$ the higher the rate coverage probability or equivalently, the smaller the SINR outage probability. That holds true essentially because smaller values of $\psi$ are associated with higher antenna gains. Finally we answer to \textbf{[Q2]} by noting that, for $\psi = 45^\circ$, user $o$ can successfully support incoming data streams from $M_{\mathcal{C}_o}$ or $m_{\mathcal{C}_o}$ at a rate greater than $\SI{4.6}{\Gbps}$ or $\SI{4.3}{\Gbps}$ with a probability of $0.8$.

\bibliographystyle{IEEEtran}
\bibliography{IEEEabrv,papers}
\end{document}